\documentclass[12pt, a4paper,hidelinks]{article}

\usepackage{graphics}
\usepackage{fullpage,epsf,graphicx, amstext, amsmath,url,tikz,hyperref,array} 
\usepackage{amssymb}
\usepackage{graphicx}
\usepackage{caption}
\usepackage{subcaption}
\usepackage{float}
\usepackage{braket}
\usepackage{cite}

\usetikzlibrary{positioning,arrows}
\usetikzlibrary{calc,backgrounds}
\usetikzlibrary{decorations.pathmorphing}
\usetikzlibrary{decorations.markings}

\tikzset{
	particle/.style={thick,draw=blue, postaction={decorate},
		decoration={markings,mark=at position .5 with {\arrow[blue]{triangle 45}}}},
	gluon/.style={decorate, draw=black,
		decoration={coil,aspect=0}},
	photon/.style={decorate, decoration={snake}},
	dot/.style={circle,fill=black,inner sep=0pt,minimum size=.2cm},
	dotg/.style={circle,fill=black!25,inner sep=0pt,minimum size=.2cm},
	dotw/.style={circle,draw=black,inner sep =0cm,minimum size=.2cm} }

\newcommand*\dd{\mathop{}\!\mathrm{d}}
\newcommand\com{{\mathbb C}}
\newcommand\cN{{\mathcal N}}
\newcommand\cZ{{\mathcal Z}}
\newcommand\Gr{\text{Gr}}
\newcommand*\diff{\mathop{}\!\mathrm{d}}

\DeclareMathOperator{\sgn}{sgn}

\begin{document}

\thispagestyle{empty}

\null\vskip-43pt \hfill
\begin{minipage}[t]{50mm}
	DCPT-18/21 \\
	\end{minipage}

\vskip 2.2truecm
\begin{center}
	\vskip 0.2truecm

	{\Large\bf
		%\titleline
		The twistor Wilson loop and the amplituhedron
	}
	\vskip 0.5truecm
	%\vfill
	
	\vskip 1truecm
	%\vfill
	{\bf   Paul Heslop and Alastair Stewart\\
	}
	
	\vskip 0.4truecm
	% \addresses
	{\it 
		Mathematics Department, Durham University,
		Science Laboratories,
		\\South Rd, Durham DH1 3LE}
	 \\
	
\end{center}

\vskip 2truecm

\centerline{\bf Abstract} % \normalsize
\medskip \noindent 
The amplituhedron provides a beautiful description of perturbative superamplitude integrands in $\cN=4$ SYM in terms of purely geometric objects, generalisations of   polytopes. 
On the other hand  the Wilson loop in supertwistor space also gives an explicit description of these  superamplitudes as a sum of planar Feynman diagrams. Each Feynman diagram can be naturally associated with a geometrical object in the same space as the amplituhedron (although not uniquely). This suggests that these geometric images of the Feynman diagrams give a tessellation of the amplituhedron. This turns out to be the case for NMHV amplitudes. We prove however that beyond NMHV this is not true. Specifically, each Feynman diagram leads to an image with a physical boundary and spurious boundaries. The spurious ones should be ``internal", matching with neighbouring diagrams. We however show that   there is no choice of geometric image of the Wilson loop Feynman diagrams which yields a geometric object without leaving unmatched spurious boundaries.

\newpage

\thispagestyle{empty}

\tableofcontents

\section{Introduction}

Scattering amplitudes in planar $\cN=4$ SYM have long been a fruitful source of new concepts and techniques in quantum field theory. 
One of the most exciting recent discoveries relates their perturbative integrands  directly to  geometric objects.  This was first noticed by Hodges \cite{0905.1473}, and was further developed by Arkani-Hamed \textit{et al} \cite{1012.6030}.  Arkani-Hamed and Trnka then interpreted the integrands as being equivalent to generalised polyhedra in positive Grassmannians called `amplituhedra'~\cite{1312.2007}.  
This has lead to a great deal of interest from both physicists and mathematicians as well as a number of generalisations~\cite{Arkani-Hamed:2013kca,Bai:2014cna,Franco:2014csa,Lam:2014jda,Arkani-Hamed:2014dca,Agarwala:2015vma,Bai:2015qoa,Ferro:2015grk,Bern:2015ple,Galloni:2016iuj,Karp:2016uax,Dennen:2016mdk,Ferro:2016zmx,Ferro:2016ptt,Eden:2017fow,Arkani-Hamed:2017vfh,Karp:2017ouj,Rao:2017fqc,An:2017tbf,Arkani-Hamed:2017fdk,Arkani-Hamed:2017mur,Galashin:2018fri,Agarwala:2018fms,Ferro:2018vpf,Rao:2018uta,Bourjaily:2018bbb}.

Although early polytope interpretations~\cite{0905.1473,1012.6030} involved considering amplitudes via twistor Wilson loop diagrams (WLDs) the amplituhedron itself instead arose from considering the BCFW method of obtaining amplitudes. However the WLDs apparently lend themselves very naturally and directly to a geometrical interpretation and  in this paper we wish to look again at the relationship between WLDs and the amplituhedron.  Previous work also examining this connection includes~\cite{Agarwala:2015vma,Eden:2017fow,Agarwala:2018fms}. In particular in~\cite{Agarwala:2018fms} it was shown that the WLDs give a very natural description of the physical boundary of the amplituhedron.
Specifically here we wish to examine whether it is possible to use WLDs to define a tessellation of the amplituhedron or more generally a tessellation of any ``good" geometrical shape, whereby ``good" means it only has a physical  boundary (corresponding to poles of the amplitude) without any spurious boundaries.  We prove that beyond NMHV this is not the case. The WLDs do not give a tessellation of the amplituhedron or any other geometrical object without remaining unmatched spurious boundaries.

Let us emphasise that we make no assumptions about positivity, or convexity or any particular specific form for this geometrical shape. Our only assumptions are that each WLD is associated with a region of amplituhedron space in such a way that the canonical form~\cite{1703.04541} of that region gives back the WLD. Since each WLD contains spurious poles which have a geometrical interpretation as spurious boundaries we then  ask if it is possible to choose these regions in such a way that all spurious boundaries glue together correctly pairwise with those of other diagrams so that the union of regions leaves no remaining unmatched spurious boundaries. This turns out to be impossible.

Many of the salient points can be illustrated in the toy model for the amplituhedron introduced in~\cite{1312.2007} consisting simply of polygons in $P^2$ with $n$ vertices $Z_i \in P^2$. The map from this polygon $X$  to the algebraic ``amplitude'' $\Omega(X)$ is made by associating  a ``canonical form'' with the geometry. This canonical form is a differential volume form with logarithmic divergences on the boundary of the polygon and no divergences inside it. Such differential forms are not easy to obtain directly~\cite{1703.04541}, but have the helpful feature that the volume form of the union of (non-overlapping) polygons gives the sum of the volume forms for each i.e.  $\Omega(X_1 \cup X_2) = \Omega(X_1)+\Omega(X_2)$. This gives a simple means of obtaining the canonical form for a polygon by triangulating it and summing the canonical forms for each triangle.

A simple way to obtain the canonical form for a triangle with vertices $Z_1,Z_2,Z_3$ is to choose coordinates $a,b$ such  that the inside of the triangle coincides with the region $a,b>0$ i.e.   $Y= a Z_1 +b Z_2 +Z_3$. Then the canonical form is simply $\dd a \dd b/(ab)$  which can then be rewritten in a co-ordinate independent way as $\langle Y \dd^2 Y \rangle \langle 1 2 3 \rangle^2/ (\langle Y 1 2  \rangle\langle Y 2 3 \rangle\langle Y  3 1 \rangle)$.
Two adjacent triangles with vertices $Z_1,Z_2,Z_3$ and $Z_1,Z_3,Z_4$ triangulate a quadrilateral with vertices $Z_1,Z_2,Z_3,Z_4$. Each individual triangle has a boundary $[Z_1Z_3]$ which is not a boundary of the quadrilateral. Such a boundary is referred to as ``spurious''. Similarly each canonical form has a corresponding log divergence when $Y$ approaches this boundary,  $ Y \rightarrow \alpha Z_1 +\beta Z_3 $.  However in the sum of the two canonical forms, $\langle 1 2 3 \rangle^2/ (\langle Y 1 2  \rangle\langle Y 2 3 \rangle\langle Y  3 1 \rangle)+\langle 1 3 4 \rangle^2/ (\langle Y 1 3 \rangle\langle Y  3 4 \rangle\langle Y 4 1  \rangle)$ the residues of the two poles cancel there and the resulting canonical form indeed only has log divergences on the boundary of the quadrilateral itself. 

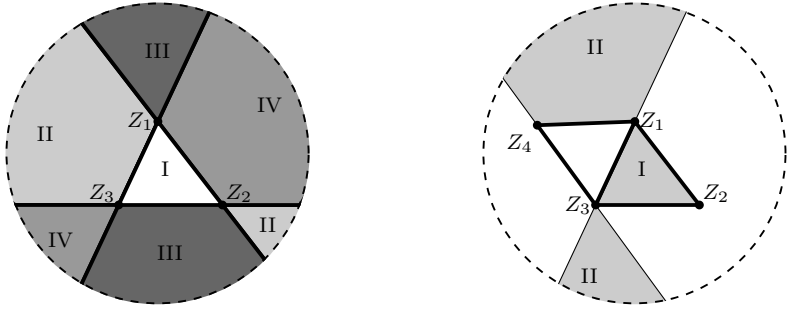
\begin{figure}[t]
	\begin{center}
\begin{tikzpicture}
	\coordinate (l1a) at draw (-20:2);
	\coordinate (l1b) at draw (200:2);
	\coordinate (l2a) at draw (120:2);
	\coordinate (l2b) at draw (-45:2);
	\coordinate (l3a) at draw (70:2);
	\coordinate (l3b) at draw (-120:2);
	\coordinate (l1l2) at (intersection of l1a--l1b and l2a--l2b);
	\coordinate (l1l3) at (intersection of l1a--l1b and l3a--l3b);
	\coordinate (l2l3) at (intersection of l3a--l3b and l2a--l2b);
		\clip (0,0) circle (2cm);
	\path[fill=black!20!white,draw=black,line width=0.5mm] (l1a) --  (l1b) -- (150:7) -- (l2a) -- (l2b) -- (-30:7) -- cycle;
	\path[fill=black!40!white,draw=black,line width=0.5mm] (l1a) --  (l1b) -- (-150:7) -- (l3b) -- (l3a) -- (45:7) -- cycle;
	\path[fill=black!60!white,draw=black,line width=0.5mm] (l3a) --  (l3b) -- (-80:7) -- (l2b) -- (l2a) -- (90:7) -- cycle;
	\path[fill=white,draw=black,line width=0.5mm] (l1l2) --  (l1l3) -- (l2l3) -- cycle;
	\draw[line width=.5mm,dashed] (0,0) circle (2cm);
	\coordinate[label=center:{$\scriptstyle \textrm{I}$}] (I) at (.1,-.2);
	\coordinate[label=center:{$\scriptstyle \textrm{II}$}] (II1) at (170:1.5);
		\coordinate[label=center:{$\scriptstyle \textrm{II}$}] (II2) at (-33:1.73);
			\coordinate[label=left:{$\scriptstyle \textrm{III}$}] (III1) at (77:1.4);
			\coordinate[label=left:{$\scriptstyle \textrm{III}$}] (III2) at (-70:1.4);
			\coordinate[label=left:{$\scriptstyle \textrm{IV}$}] (IV1) at (20:1.9);
			\coordinate[label=left:{$\scriptstyle \textrm{IV}$}] (IV2) at (230:1.5);
			\node [fill,circle,scale=0.3,label={[label distance=-2mm]45:$\scriptstyle Z_2$}] at (l1l2) {};
			\node [fill,circle,scale=0.3,label={[label distance=-2mm]135:$\scriptstyle Z_3$}] at (l1l3) {};
			\node [fill,circle,scale=0.3,label={[label distance=-1.5mm]180:$\scriptstyle Z_1$}] at (l2l3) {};
			\end{tikzpicture}
\hspace{2cm}
\begin{tikzpicture}
\coordinate (l1a) at draw (-20:2);
\coordinate (l1b) at draw (200:2);
\coordinate (l2a) at draw (120:2);
\coordinate (l2b) at draw (-45:2);
\coordinate (l3a) at draw (70:2);
\coordinate (l3b) at draw (-120:2);
\coordinate (l4a) at draw (150:2);
\coordinate (l5a) at draw (170:2);
\coordinate (low1) at draw (-10,-10);
\coordinate (low2) at draw (10,-10);
\coordinate (l1l2) at (intersection of l1a--l1b and l2a--l2b);
\coordinate (l1l3) at (intersection of l1a--l1b and l3a--l3b);
\coordinate (l2l3) at (intersection of l3a--l3b and l2a--l2b);
\coordinate (l4b) at (intersection of l4a--l1l3 and low1--low2);
\coordinate (l4l5) at (intersection of l4a--l4b and l5a--l2l3);
\clip (0,0) circle (2cm);
\path[fill=white!80!black,draw=black] (l1l3) --  (l3b) -- (-90:7) -- (l4b) --  cycle;
\path[fill=white!80!black,draw=black] (l2l3) --  (l4l5) -- (l4a) -- (90:7) -- (l3a) -- cycle;
\path[fill=white!80!black,draw=black,line width=.5mm] (l1l2) --  (l1l3) -- (l2l3) -- cycle;
\path[fill=white,draw=black,line width=.5mm] (l2l3) --  (l4l5) -- (l1l3) -- cycle;
\draw[line width=.5mm,dashed] (0,0) circle (2cm);
	\coordinate[label=center:{$\scriptstyle \textrm{I}$}] (I) at (.1,-.2);
\coordinate[label=center:{$\scriptstyle \textrm{II}$}] (II1) at (110:1.5);
\coordinate[label=center:{$\scriptstyle \textrm{II}$}] (II2) at (-110:1.73);
	\node [fill,circle,scale=0.3,label={[label distance=-1.5mm]0:$\scriptstyle Z_1$}] at (l2l3) {};
			\node [fill,circle,scale=0.3,label={[label distance=-2mm]45:$\scriptstyle Z_2$}] at (l1l2) {};
\node [fill,circle,scale=0.3,label={[label distance=-1.5mm]180:$\scriptstyle Z_3$}] at (l1l3) {};
\node [fill,circle,scale=0.3,label={[label distance=-1.5mm]190:$\scriptstyle Z_4$}] at (l4l5) {};
\end{tikzpicture}
\end{center}
\caption{Figures illustrating polygons in $P^2$ represented as a disc where opposite points of the disc are identified. In Figure a) we illustrate the fact that there are four triangles $\mathrm{I,II,III,IV}$ all of which have the same three vertices $Z_1,Z_2,Z_3$ and all having the same canonical form $\langle 1 2 3 \rangle^2/ (\langle Y 1 2  \rangle\langle Y 2 3 \rangle\langle Y  3 1 \rangle)$. In Figure b) we see a region (shaded area) which has the same canonical form as the quadrilateral $[Z_1Z_2Z_3Z_4]$, $\langle 1 2 3 \rangle^2/ (\langle Y 1 2  \rangle\langle Y 2 3 \rangle\langle Y  3 1 \rangle)+\langle 1 3 4 \rangle^2/ (\langle Y 1 3 \rangle\langle Y  3 4 \rangle\langle Y 4 1  \rangle)$ but which does not represent a good geometrical region as it has spurious boundaries.  }
\label{P2examples}
\end{figure}

Although there is a unique canonical form associated to a polygon, the reverse is not true. For example, given the canonical form for the triangle with vertices $Z_1,Z_2,Z_3$, $\langle Y \dd^2 Y \rangle \langle 1 2 3 \rangle^2/ (\langle Y 1 2  \rangle\langle Y 2 3 \rangle\langle Y  3 1 \rangle)$, there are {\em four inequivalent } triangles in $P^2$  with this canonical form. These are given by the set $\{Y: Y= a Z_1 +b Z_2 +Z_3\}$ for the four choices $(a,b>0)$, $(a>0,b<0)$, $(a<0,b>0)$ or $(a<0,b<0)$. These are simply the four inequivalent  triangles in $P^2$ with vertices $Z_1,Z_2,Z_3$ (see Figure~\ref{P2examples}a).

So the  geometry associated with a given canonical form is not unique but only defined  up to sign choices. If on the other hand we are given a canonical form, written as a sum of terms each containing spurious poles that cancel in the sum (which as we will see is precisely what WLDs give us), then the assigning of a geometrical region to each term (i.e. the choice of signs) can not be done independently for each term: the cancelling of spurious poles should correspond geometrically to a matching of the corresponding spurious boundaries
(in Figure~\ref{P2examples}b we see a simple example of a region with the same canonical form as the quadrilateral $[Z_1Z_2Z_3Z_4]$ but with left over spurious boundaries).

\begin{figure}[H]
	\centering
	\includegraphics[scale=0.6]{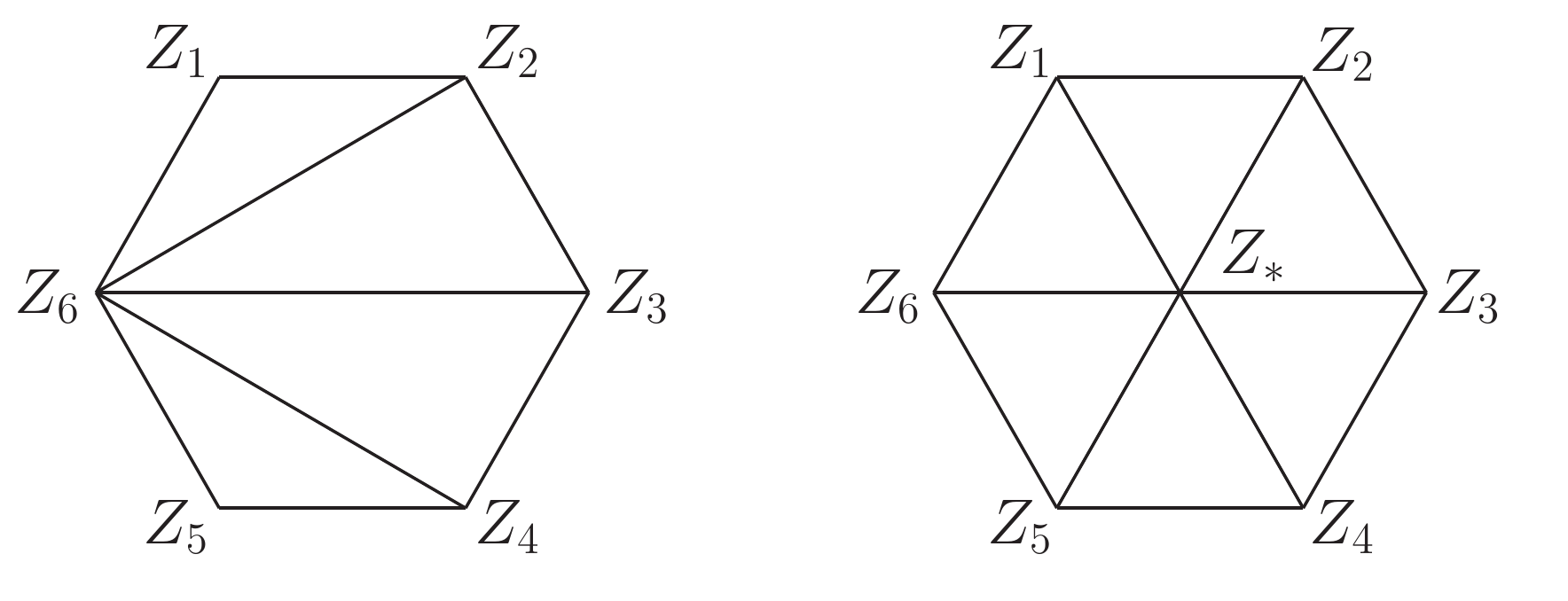} 
	\caption{Two possibilities for triangulating a polygon. BCFW give a generalisation of the first whereas WLDs give a generalisation of the second (for NMHV).}
	\label{tessellation}
\end{figure}

There are two natural ways to triangulate a polygon illustrated in Figure~\ref{tessellation}. 
BCFW recursion for tree-level NMHV diagrams gives the natural (higher dimensional) analogue of the first way,  triangulating to one of the vertices. 

Remarkably WLDs for the planar N${}^k$MHV amplitude/ Wilson loop split the amplitude into well-defined pieces, each of which naturally yields a volume form on the space on which the amplituhedron lies, the Grassmannian  $\Gr(k,4{+}k)$. Each volume form has physical poles and spurious poles the latter of which all cancel in the sum over diagrams. The physical poles of the WLDs correspond to  the physical boundary of the amplituhedron~\cite{Agarwala:2018fms}. 
This therefore strongly suggests that the WLDs should correspond to be a tessellation of the amplituhedron. The canonical forms of each tile corresponding to WLDs. Note that if this were the case the  WLDs would then give a very explicit tessellation of the amplituhedron. 

In the NMHV tree-level case this intuition indeed turns out to be correct: each WLD can be straightforwardly associated with a tile in the tessellation of the amplituhedron. Indeed NMHV Twistor Wilson loop Feynman diagrams (WLDs) naturally give a higher dimensional analogue of  the second way of tessellating polygons, introducing an additional vertex $Z_*$ and triangulating to that, Figure~\ref{tessellation}b. 

In this paper we however prove that, for higher NMHV degree this is not the case. More concretely we prove that there do not exist a set of tiles whose canonical forms correspond to WLDs and which glue together to form a geometry without spurious boundaries. The WLDs can therefore not be associated with a tessellation of the amplituhedron or indeed any geometry whose boundary corresponds to only the physical poles of the amplitude.

{\bf Note added: The paper~\cite{1807.05397} by Susama Agarwala and Cameron Marcott dealing with the same problem as this paper was posted on the same day.}

\section{WLDs and volume forms}

\subsection{WLDs}
\label{wlds}
Here, we provide a brief description of planar Wilson loops in ${\mathcal N}=4$ Super Yang Mills in super twistor space and define the WLDs that arise. We do not derive these here, for their derivation  see~\cite{Mason:2010yk,Chicherin:2016fac,Chicherin:2016ybl}.

\begin{figure}[H]
  	\begin{center}
  		\includegraphics[scale=0.55]{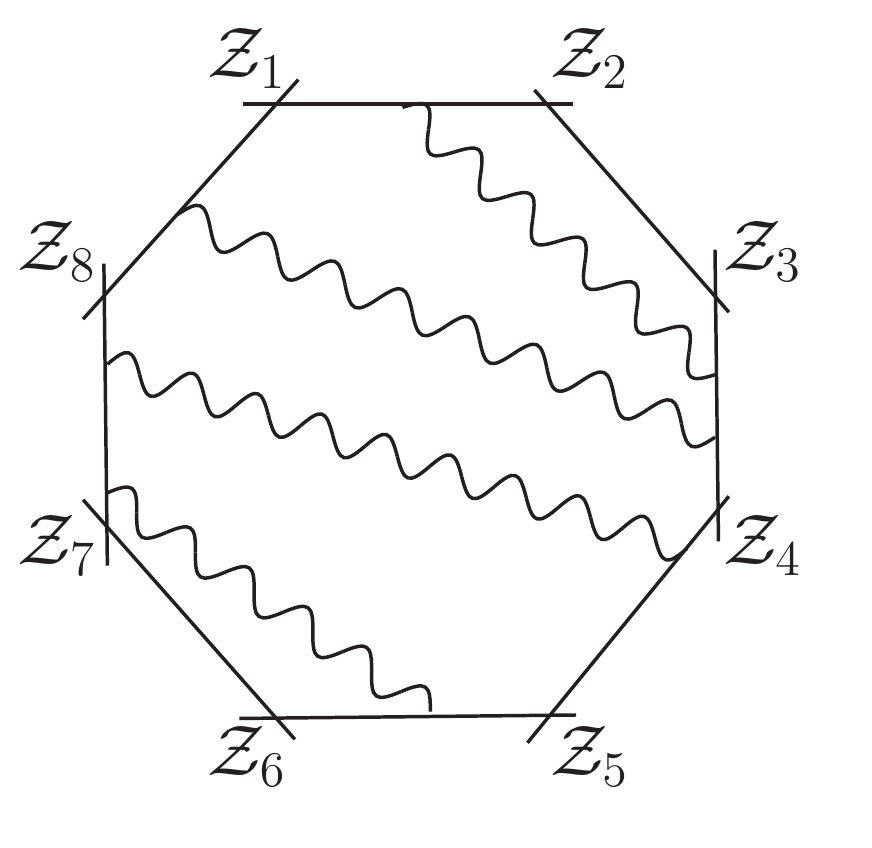}
  		\caption{Example of a Wilson loop diagram which contributes to the $8$-point N${}^4$MHV amplitude.}
  		\label{fig:WLDExample}
  	\end{center}
  \end{figure}

 The WLDs we are discussing here are simply the Feynman diagrams describing a polygonal holomorphic Wilson-loop in super twistor space  with vertices being the super twistors $\mathcal{Z}_1 \dots \mathcal{Z}_n \in {\mathbb C}^{4|4}$. In the planar theory this is  equivalent, via the Wilson loop / amplitude duality~\cite{Alday:2007hr,Drummond:2007aua,Brandhuber:2007yx}, to $n$-point superamplitudes. 
  At tree level the Feynman diagrams consist  simply of  propagators whose two ends lie on the Wilson loop contour. In the planar theory  diagrams are only valid if we can draw all the propagators inside the Wilson loop without crossing.  The N${}^k$MHV Wilson loop is the sum over all such diagrams involving $k$ propagators (see Figure~\ref{fig:WLDExample} for an example of a diagram contributing to  8-point N${}^4$MHV).
  
To each propagator from edge $[\cZ_i\cZ_{i+1}]$ to $[\cZ_j\cZ_{j+1}]$ we assign the $(4|4)$ delta function:
\begin{align}
\begin{tikzpicture}
\def\width{5cm}
\def\height{1.5cm}
\coordinate[label=left:{$\cZ_{i+1}$}](zip1);
\coordinate[below=\height of zip1,label=left:{$\cZ_i$}](zi);
\coordinate[below=\height/2 of zip1,label=above right:{$\scriptstyle b$},label=below right:{$\scriptstyle a$}](pl);
\coordinate[right=\width of pl,label=above left:{$\scriptstyle c$},label=below left:{$\scriptstyle d$}](pr);
\coordinate[right=\width of zip1,label=right:{$\cZ_j$}](zj);
\coordinate[right=\width of zi,label=right:{$\cZ_{j+1}$}](zjp1);
\draw[thick] (zi) -- (zip1); 
\draw[thick] (zj) -- (zjp1); 
\draw[photon] (pl) -- (pr);
\end{tikzpicture}
\raisebox{1cm}{$\  \displaystyle =\ \delta^{4|4}(a \cZ_i {+}b \cZ_{i{+}1} {+}c \cZ_{j} {+}d \cZ_{j+1} {+}\cZ_*)$  }
\label{proprule}
\end{align}
We then integrate over the complex integration variables associated with each end of the propagator with a measure determined by all the propagators ending on the same edge
\begin{align}
\begin{tikzpicture}
\def\width{5cm}
\def\height{1cm}
\coordinate(zi);
\coordinate[right=\width of zi](zip1);
\coordinate[right=\width/10 of zi,label=below left:{$\scriptstyle a_1$},label=below right:{$\scriptstyle b_1$}](p1);
\coordinate[right=\width/5 of p1,label=below left:{$\scriptstyle a_2$},label=below right:{$\scriptstyle b_2$}](p2);
\coordinate[right=1.7*\width/5 of p2,label=below left:{$\scriptstyle a_{m\!-\!1}$},label=below right:{$\scriptstyle b_{m\!-\!1}$}](p3);
\coordinate[right=1.3*\width/5 of p3,label=below left:{$\scriptstyle a_m$},label=below right:{$\scriptstyle b_m$}](p4);
\coordinate[below=\height of p1](q1);
\coordinate[below=\height of p2](q2);
\coordinate[below=\height of p3](q3);
\coordinate[below=\height of p4](q4);
\coordinate[below right=\height of p1,label=right:{$\ \ \ \ \  \dots$}](dots);
\draw[thick] (zi) -- (zip1); 
\draw[photon] (p1) -- (q1);
\draw[photon] (p2) -- (q2);
\draw[photon] (p3) -- (q3);
\draw[photon] (p4) -- (q4);
\end{tikzpicture}
\raisebox{.5cm}{$\  \displaystyle =\ \int \frac{\dd a_1 \dd b_1 \dots \dd a_m \dd b_m}{b_1 (a_1b_2{-}b_1a_2) \dots (a_{m-1}b_m{-}b_{m-1}a_m)a_m} $  }
\label{vertrule}
\end{align}

\begin{figure}[H]
$	\begin{array}{m{8cm}m{8cm}}
		\includegraphics[scale=0.5]{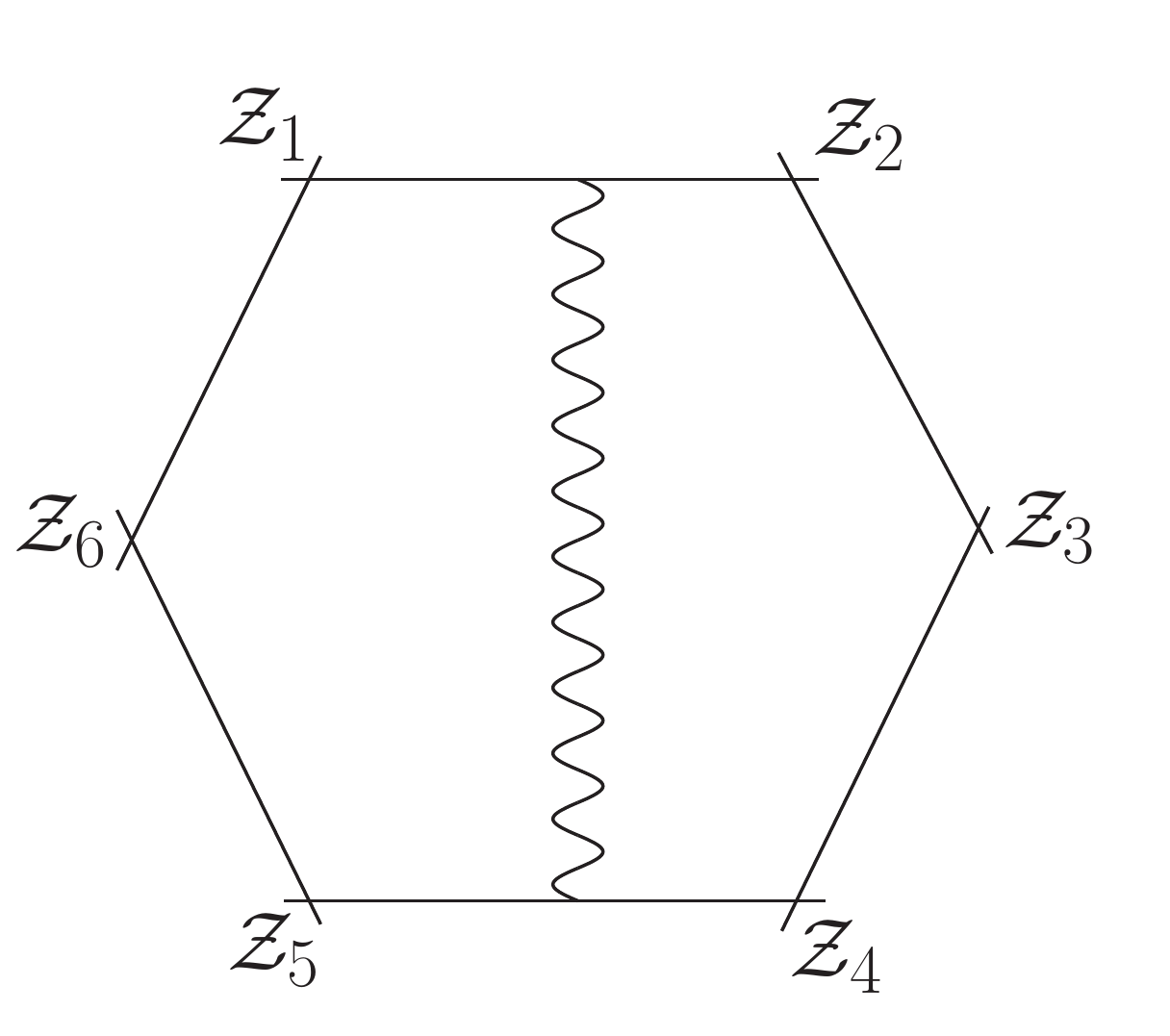}&
		\includegraphics[scale=0.5]{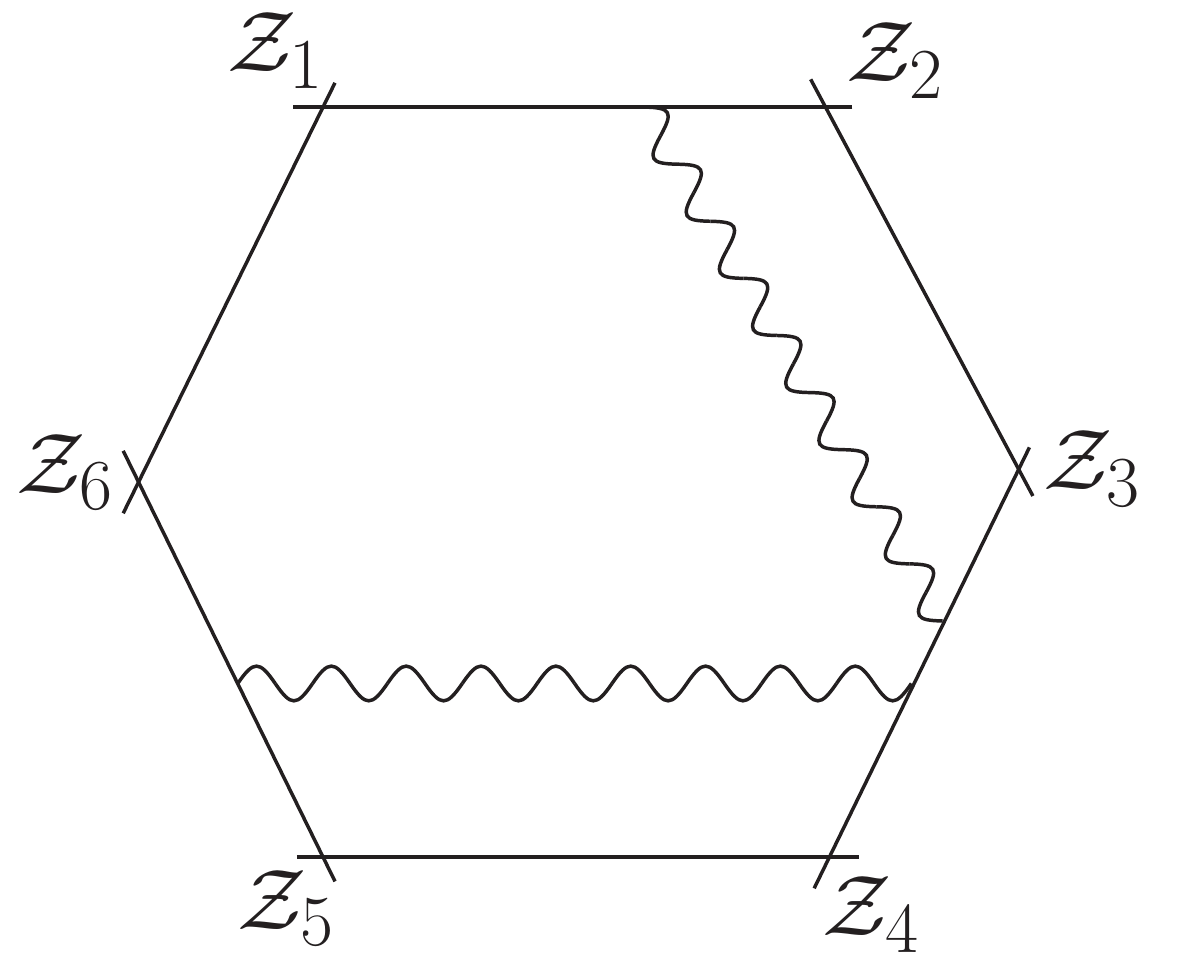}\\
		$\int \frac{\dd a\dd b\dd c\dd d}{abcd} {\delta}^{4|4}\left(  a\cZ_1 {+} b \cZ_2 {+} c\cZ_4 {+} d\cZ_5  {+}\cZ_*\right)$&
		$
		\begin{array}{l}
		\int \frac{\dd a_1\dd b_1\dd c_1\dd  d_1 \dd f_1\dd g_1\dd h_1}{a_1b_1g_1h_1e_1(c_1f_1-d_1e_1)d_1}\times\\
		\times\delta^{(4|4)}\left(  a_1\cZ_1 {+} b_1 \cZ_2 {+} c_1\cZ_3 {+} d_1\cZ_4  {+}\cZ_*\right)\\
		\times \delta^{(4|4)}\left(  e_1\cZ_3 {+} f_1 \cZ_4 {+} g_1\cZ_5 {+} h_1\cZ_6  {+}\cZ_*\right)
		\end{array}$
		\end{array}
		$
		\caption{Examples of Feynman diagrams in twistor space that contribute to the $6$-point NMHV / N${}^2$MHV amplitude with their corresponding expressions following the rules given.}
		\label{fig:NMHVExample}
\end{figure}

In Figure~\ref{fig:NMHVExample} we illustrate these rules with two examples firstly an example diagram contributing to the NMHV six-point amplitude and secondly one contributing to the  N${}^2$MHV six-point amplitude.

\subsection{Amplituhedron volume forms from WLDs}

The WLDs are originally defined  in supertwistor space, ${\mathbb C}^{4|4}$, but have a very direct interpretation  as volume forms in the Grassmannian of $k$-planes in ${\mathbb C}^{4+k}$, $\Gr(k,4{+}k)$ or ``amplituhedron space". 

Essentially the integration variables and delta functions of the WLDs define coordinates in amplituhedron space, and the measure gives the volume form written in terms of these coordinates. 
All N${}^k$MHV WLDs have the general form, following from the description in the previous subsection
\begin{align}
	\int \Omega_{4k}(a_i) \delta^{k \times (4|4)}(C(a_i).\cZ) \qquad \text{[WLD]}
\end{align} 
where $a_i$ are the $4k$ coordinates (4 for each of the $k$ propagators),  $\Omega_{4k}(a_i)$ is the integration measure (a $4k$-form obtained as a product of terms of the form~\eqref{vertrule}) and $\delta^{4k|4k}(C(a_i).\cZ)$ are the $k$ delta functions, one for each propagator as in~\eqref{proprule},  written as a $k\times (n{+}1)$ matrix $C(a_i)$ acting on the external supertwistors $\cZ$, themselves viewed as an $(n{+}1) \times (4|4)$ matrix.
The corresponding volume form in $\Gr(k,k{+}4)$ is then simply the measure $\Omega_{4k}(a_i)$ where the coordinates are now reinterpreted as co-ordinates in $\Gr(k,k{+}4)$ via the map
\begin{align}
	\Omega(Y) = \Omega_{4k}(a_i)  \qquad     Y= C(a_i).Z \in \Gr(k,k+4) \label{volform}
\end{align}
and $Z$ is here an $(n{+}1)\times (4+k)$ matrix, the external $\cZ$s converted to $4+k$ dimensional bosonised supertwistors in the standard way described in~\cite{1312.2007}.

We illustrate this using  the two examples of Figure~\ref{fig:NMHVExample}.
For the NMHV example diagram of Figure~\ref{fig:NMHVExample}a
we read off the volume form:
\begin{align}
\begin{array}{cl}
	\int \frac{\dd a\dd b\dd c\dd d}{abcd} \,{\delta}^{4|4}\left(  a\cZ_1 {+} b \cZ_2 {+} c\cZ_4 {+} d\cZ_5{+}\cZ_* \right)   &[\text{WLD}] \\\downarrow \\
	\Omega=\frac{\dd a\dd b\dd c\dd d}{abcd}  \qquad Y= aZ_1 {+} b Z_2 {+} cZ_4 {+} dZ_5{+}Z_*\in \com^{5} &\text{[Amplituhedron Volume form]}\label{NMHVdiag}
\end{array}
\end{align}
This volume form can be covariantised to be written in a coordinate independent way as
\begin{align}
	\frac{\langle Y\dd^4Y \rangle \langle Z_1 Z_2 Z_4 Z_5 Z_*  \rangle^4}{\langle Y Z_1 Z_2 Z_4 Z_5\rangle \langle Y Z_2 Z_4 Z_5 Z_* \rangle\langle Y  Z_4 Z_5  Z_* Z_1 \rangle\langle Y  Z_5 Z_* Z_1 Z_2 \rangle\langle Y Z_*Z_1 Z_2 Z_4 \rangle} \ ,\label{NMHVdiagcov}
\end{align}
where the angle brackets denote $5\times 5$ determinants.

For the second  N${}^2$MHV example diagram of Figure~\ref{fig:NMHVExample}b we get 
\begin{align}
\begin{array}{cl}
\int \frac{\dd a_1\dd b_1\dd c_1\dd  d_1 \dd f_1\dd g_1\dd h_1}{a_1b_1g_1h_1e_1(c_1f_1-d_1e_1)d_1}\delta^{(8|8)}\left( C_1\cdot\mathcal{Z}\right)   &[\text{WLD}] \\\downarrow \\
\Omega=\frac{\dd a_1\dd b_1\dd c_1\dd  d_1 \dd f_1\dd g_1\dd h_1}{a_1b_1g_1h_1e_1(c_1f_1-d_1e_1)d_1}  \qquad Y= C_1.Z\in \text{Gr}(2,6) &\text{[Amplituhedron Volume form]}
\end{array}\label{volformex}
\end{align}
where $\cZ= (\cZ_1,\cZ_2, \dots \cZ_6,\cZ_*)^T$  are the external supertwistors (together with $\cZ_*$) viewed as a $7 \times (4|4)$ matrix,  
\begin{align}
	C_1= 
	\begin{pmatrix} 
	a_1 & b_1 & c_1 & d_1  & 0 & 0 & 1 \\
	0 & 0 & e_1 & f_1 & g_1 & h_1 & 1 
	\end{pmatrix} \in \text{Gr}(2,7)
\end{align}
and similarly $Z= (Z_1, \dots Z_6,Z_*)^T$  are the external bosonised supertwistors (with $Z_*$) viewed as a $7 \times 6$ matrix.

\section{NMHV amplituhedron from WLDs}

Let us first consider the NMHV case. Here the WLDs {\em do} give a good tessellation of the amplituhedron. Indeed WLDs were involved in the original polytope interpretation of amplitudes~\cite{0905.1473,1012.6030}.

The twistor  Wilson loop description of the $n$-point NMHV amplitude is simply a sum over all diagrams consisting of a single propagator attached to any two edges of the polygon.
Written as a volume form in $\Gr(1,5)$ (amplituhedron space) the WLD  corresponding to a propagator from edge $[Z_iZ_{i+1}]$ to edge $[Z_jZ_{j+1}]$ is (see~\eqref{NMHVdiag})
\begin{align}
	\Omega=\frac{\dd a\dd b\dd c\dd d}{abcd}  \qquad Y= aZ_i {+} b Z_{i+1} {+} cZ_j {+} dZ_{j+1}{+}Z_*\in \com^{5}\label{tileform}
\end{align}
which written in a coordinate independent form is~\eqref{NMHVdiagcov}
\begin{align}
	\frac{\langle Y\dd^4Y \rangle \langle Z_i Z_{i+1} Z_j Z_{j+1} Z_*  \rangle^4}{\langle Y Z_i Z_{i+1} Z_j Z_{j+1}\rangle \langle Y Z_{i+1} Z_j Z_{j+1} Z_* \rangle\langle Y  Z_j Z_{j+1}  Z_* Z_i \rangle\langle Y  Z_{j+1} Z_* Z_i Z_{i+1} \rangle\langle Y Z_*Z_i Z_{i+1} Z_j \rangle}\ .\label{covtileform}
\end{align}
So the full NMHV amplitude is thus 
\begin{align}
	 &\Omega={\langle Y\dd^4Y \rangle }\notag\\
\times 	 &\sum_{i,j}\frac{\langle Z_i Z_{i+1} Z_j Z_{j+1} Z_*  \rangle^4}{\langle Y Z_i Z_{i+1} Z_j Z_{j+1}\rangle \langle Y Z_{i+1} Z_j Z_{j+1} Z_* \rangle\langle Y  Z_j Z_{j+1}  Z_* Z_i \rangle\langle Y  Z_{j+1} Z_* Z_i Z_{i+1} \rangle\langle Y Z_*Z_i Z_{i+1} Z_j \rangle}\ .\label{NMHVamp}
\end{align}

\begin{figure}[H]
	\begin{center}
		\includegraphics[scale=0.45]{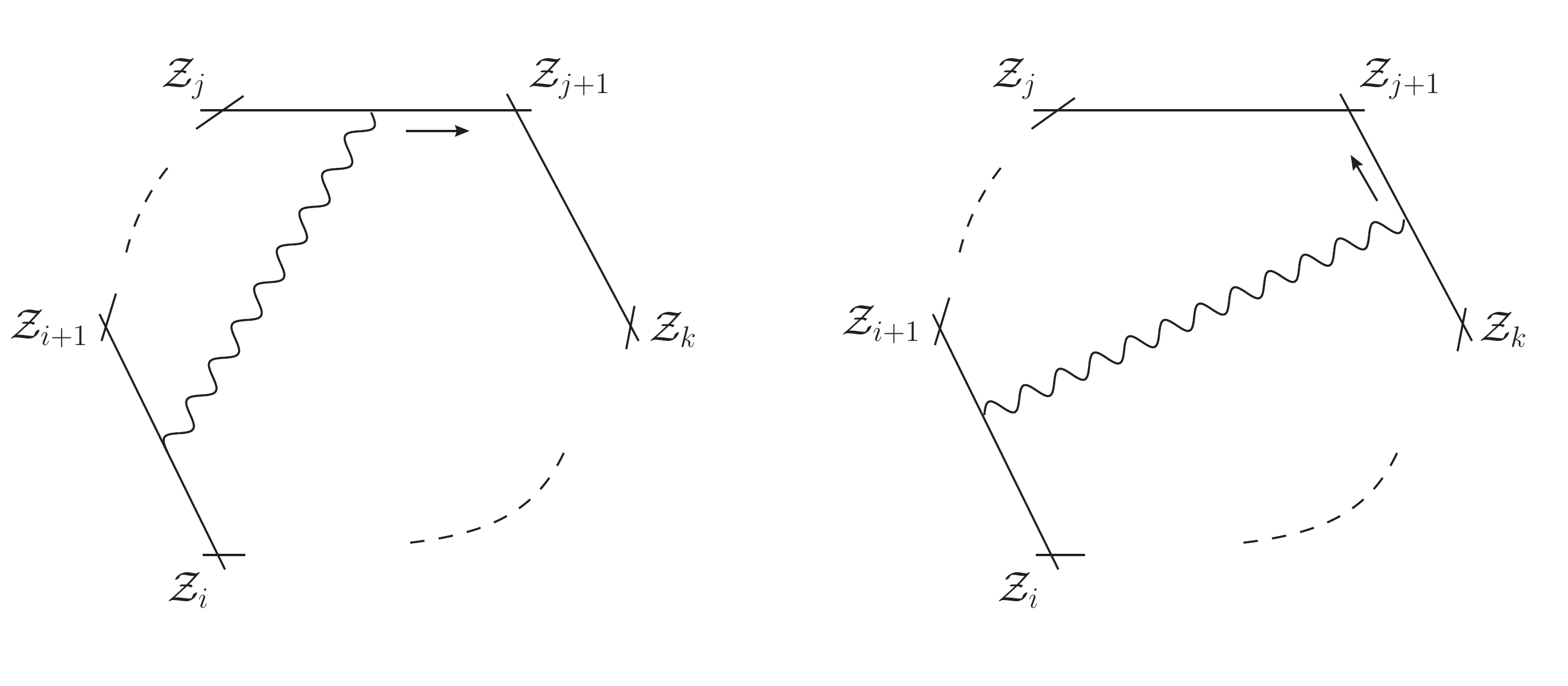}
		\caption{Spurious poles occur when the propagator end reaches the vertex. It is cancelled by the adjacent diagram. Imposing that this cancellation has a corresponding geometric meaning as a matching of spurious boundaries imposes a correlation between the  sign choices for the geometric image of the two diagrams. }
		\label{fig:MovingPropsExample}
	\end{center}
\end{figure}

It is clear from~\eqref{tileform} that the spurious poles for each WLD arise when any one of  $a,b,c,d \rightarrow 0$.%
\footnote{A fifth  pole occurs when all $a,b,c,d \rightarrow \infty$ simultaneously. This is a physical pole which does not cancel in the sum over diagrams.} 
In terms of the WLD we view this as one end of the propagator approaching a vertex. Then this spurious pole cancels with the spurious pole of a neighbouring diagram where the end of the propagator approaches the  same vertex from the other side see Figure~\ref{fig:MovingPropsExample}.

There is then  a natural geometrical interpretation of~\eqref{NMHVamp} as a union of tiles, each giving one of the above terms as its canonical form. This is
\begin{align}
	\bigcup_{i,j} \{Y=a Z_i{+}b Z_{i{+}1} {+}c Z_j {+}d Z_{j{+}1}{+}Z_*\,;\,a,b,c,d\geq 0\}\subset  {\text{Gr}}(1,5)\ .\label{NMHVtes}
\end{align}
Note here that we are using the same variables $(a,b,c,d)$ as given to us by the WLDs to describe the geometric region in question. However, whereas  for the WLDs the integration is over complex space, here the variables are restricted to a subspace of the real line.

If the $Z_i$ are convex ($\langle Z_i Z_j Z_kZ_l Z_m\rangle > 0 $ for all cyclically ordered $Z_i,Z_j,Z_k,Z_l,Z_m$)  this provides a tessellation of the amplituhedron as defined in~\cite{1312.2007}. Indeed this is analogous to the tessellation of the polygon in the toy model depicted in Figure~\ref{tessellation}b. But note that this defines a good geometrical region (i.e. one without spurious boundaries) even for non convex choices of external $Z_i$.

At this point it is interesting to ask how unique this region is. Are there any other ways of defining tiles whose canonical forms give the WLDs, and  which would glue together to yield a geometry without spurious boundaries?

As illustrated for  the toy model in Figure~\ref{P2examples}, any choice of signs for the variables $a,b,c,d$ in each tile would give a canonical form of the corresponding WLD. However if  we choose arbitrary sign choices for each diagram differently, the spurious boundaries will not glue together properly, even though the spurious poles of the corresponding canonical forms would cancel (recall  Figure~\ref{P2examples}b  for an illustration of this sort of phenomenon for the toy model). 
Let us  then consider a particular tile corresponding to the WLD with a propagator from edge $[Z_i Z_{i+1}]$ to edge $[Z_jZ_{j+1}]$. The most general geometry giving this  canonical form~(\ref{tileform},\ref{covtileform}) is
\begin{align}
 \{Y= a s_iZ_i{+}b s_{i+1}Z_{i{+}1} {+}c s_jZ_j {+}d s_{j+1}Z_{j{+}1}{+}Z_*\,:\,a,b,c,d\geq 0\}
\end{align}
where $s_i,s_{i+1},s_j,s_{j+1}=\pm1$ are four arbitrary sign choices.
The spurious poles are here seen as spurious boundaries arising when any one of the four coordinates $a,b,c,d \rightarrow 0 $  (whereas a fifth, physical boundary occurs when they  all simultaneously $a,b,c,d\rightarrow \infty$).
Let us focus on the spurious boundary when $a\rightarrow 0$. This must match the boundary when $b\rightarrow 0$ of  the adjacent diagram with propagator from edge $[Z_{i+1} Z_{i+2}]$ to edge $[Z_jZ_{j+1}]$ (which we also define with arbitrary signs $s'_{i+1},s'_{i+2},s'_j,s'_{j+1}=\pm1$):
\begin{align}
\begin{array}{c}
\{Y= a s_iZ_i{+}b s_{i+1}Z_{i{+}1} {+}c s_jZ_j {+}d s_{j+1}Z_{j{+}1}{+}Z_*\,:\,a=0,b,c,d\geq 0\} \\=\\ \{Y=as'_{i+1} Z_{i+1}{+}bs'_{i+2} Z_{i{+}2} {+}c s'_jZ_j {+}d s'_{j+1}Z_{j{+}1}{+}Z_*\,:\,b=0,\ a,c,d\geq 0\}
	\end{array}\label{spboarg}
\end{align}
This mimics the corresponding discussion of cancellation of spurious poles in Figure~\ref{fig:MovingPropsExample} and associated discussion. Except now the geometrical matching imposes consistency conditions on the sign choices of the two tiles.
For these spurious boundaries to match we clearly require
\begin{align}
	s_{i+1}=s'_{i+1}, \quad s_{j}=s'_{j}, \quad s_{j+1}=s'_{j+1}.
\end{align}
Thus the signs associated with each vertex for different diagrams must be the same. Clearly a similar mechanism applies for matching boundaries when $c$ or $d \rightarrow 0$. 

From this discussion one can see then that the most general geometry without spurious boundaries is obtained by assigning a fixed sign, $s_i=\pm 1$, to each vertex $Z_i$. So the region
\begin{align}
\bigcup_{i,j} \{Y= a s_iZ_i{+}b s_{i+1}Z_{i{+}1} {+}c s_jZ_j {+}d s_{j+1}Z_{j{+}1}{+}Z_*\,;\,a,b,c,d\geq 0\}\subset  {\text{Gr}}(1,5)\ \label{NMHVtes2}
\end{align}
is the most general geometry matching the  WLDs and without spurious boundaries.%
\footnote{One might think a more general possibility could be to have two sets of fixed signs, one for each end of the propagator. However on starting from a diagram it is possible to eventually reach the  same diagram with the ends of the propagator reversed, by matching spurious boundaries with consecutive diagrams as you go. This reversed propagator has to correspond to the same geometrical region as the original and so the two sets of signs must in fact be equal to each other.}
This is true for any choice of signs $s_i$. This is equivalent to simply considering the original amplituhedron with all positive signs but flipping the  sign of the external $Z$'s. At most one choice of signs for the $Z$s can correspond to a convex shape.
	 
For completeness we should also consider 
a special case of the spurious poles / boundaries cancellation which occurs when the propagator lies between next-to-adjacent edges, i.e. between edge $[Z_iZ_{i+1}]$ and edge $[Z_{i+2}Z_{i+3}]$.
The spurious boundary when $a=0$ of this diagram at first sight looks like it is not present (propagators between adjacent edges are not allowed; they give vanishing results). 
Instead it  matches with $d=0$ of the diagram with propagator between edge $[Z_{i+1}Z_{i+2}]$ and edge $[Z_{i+3}Z_{i+4}]$
\begin{align}
\begin{array}{c}
\{Y= a s_iZ_i{+}b s_{i+1}Z_{i{+}1} {+}c s_{i+2}Z_{i+2} {+}d s_{i+3}Z_{i+3}{+}Z_*\,;\,a=0,b,c,d\geq 0\} \\=\\ \{Y=as_{i+1} Z_{i+1}{+}bs_{i+2} Z_{i{+}2} {+}c s_{i+3}Z_{i+3} {+}d s_{i+4}Z_{i+4}{+}Z_*\,;\,d=0,\ a,b,c\geq 0\}\ .
\end{array}\label{spboarg2}
\end{align}
We see that the spurious boundaries indeed match  for this special case too  even for the  general choice of signs.

\section{$N^2$MHV}

Having considered NMHV WLDs and shown how to obtain a ``good" geometry from them (in many inequivalent  ways)   we now consider the same problem for higher MHV degree.
We will prove that beyond NMHV the WLDs cannot in fact be glued together to form a geometry without spurious boundaries.  To prove  this, it is enough to show that there is no set of sign choices for the coordinates that is consistent with the matching of spurious boundaries.  In order to illustrate the argument we focus on  the case of $n=6$ below, however the  argument applies to all $n$.

\subsection{Cancellation of spurious poles in $N^2$MHV WLDs}

\begin{figure}[H]
	\centering
	\includegraphics[scale=0.4]{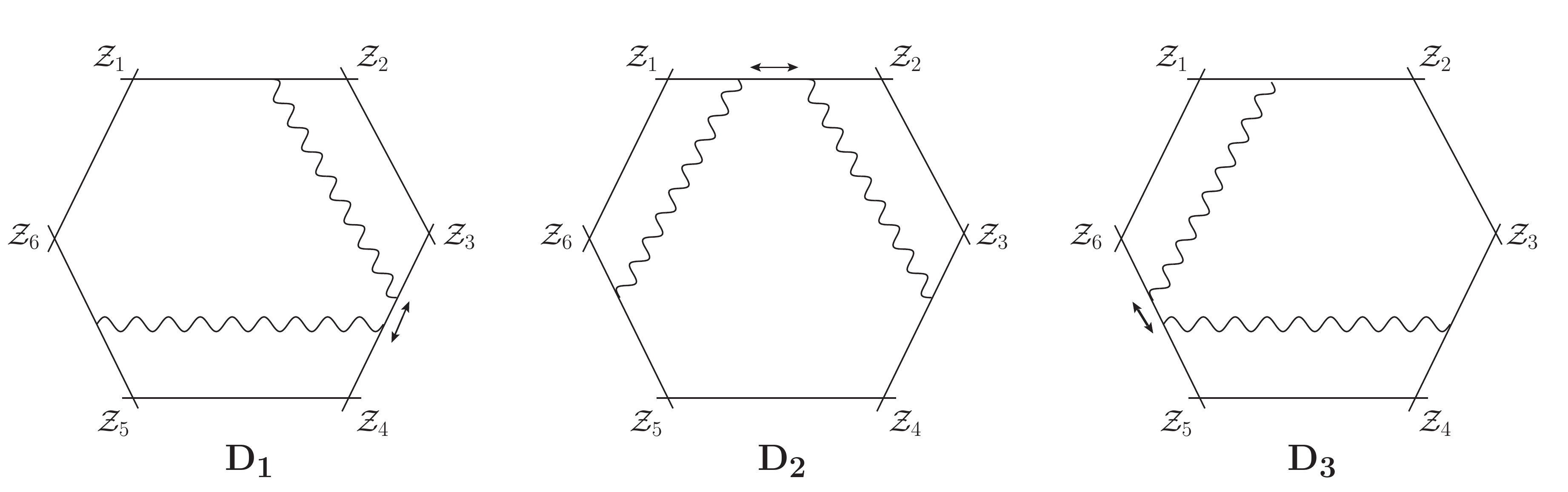}
	\caption{Three diagrams each having a new type of spurious pole occurring when the propagator ends touch. In the sum of the three diagrams however this pole cancels. Note that although this is drawn at six points for definiteness the cancellation only depends on the three sides taking  part and can be directly repeated at $n$ points. }
	\label{fig:threediag}
\end{figure}

Before considering the geometric image as spurious boundaries we consider the algebraic cancellation of spurious poles for N${}^2$MHV diagrams. The discussion of spurious poles considered in the previous section, arising when the ends of propagators approach vertices (see Figure~\ref{fig:MovingPropsExample}) goes through in the same way for any MHV degree. However beyond NMHV a new type of spurious pole occurs  in the integrals of WLDs. Since now we have two or more propagators, there exists the possibility that the ends of two different propagators can meet each other on an edge. This produces a pole in the WLD.  There is an interesting three-way cancellation of this type of spurious pole between three related diagrams (see~\cite{Chicherin:2014uca,Chicherin:2016ybl} for previous  work also describing this mechanism).  An example set of diagrams is shown in Figure~\ref{fig:threediag}.

Using the rules from section~\ref{wlds}, the integrals associated with the diagrams under consideration are
\begin{equation}
\label{eq:C1Integral}
\mathcal{I}(D_1)= \int \frac{\diff a_1 \diff b_1 \diff c_1 \diff d_1 \diff e_1 \diff f_1 \diff g_1 \diff h_1}{a_1b_1g_1h_1e_1(c_1f_1-d_1e_1)d_1}\delta^{(8|8)}\left( C_1\cdot\mathcal{Z}\right)
\end{equation}
\begin{equation}
\label{eq:C2Integral}
\mathcal{I}(D_2)= \int \frac{\diff a_2 \diff b_2 \diff c_2 \diff d_2 \diff e_2 \diff f_2 \diff g_2 \diff h_2}{c_2d_2g_2h_2b_2(a_2f_2-b_2e_2)e_2}\delta^{(8|8)}\left( C_2\cdot\mathcal{Z}\right)
\end{equation}
and
\begin{equation}
\label{eq:C3Integral}
\mathcal{I}(D_3)= \int \frac{\diff a_3 \diff b_3 \diff c_3 \diff d_3 \diff e_3 \diff f_3 \diff g_3 \diff h_3}{a_3b_3e_3f_3c_3(d_3g_3-h_3c_3)h_3}\delta^{(8|8)}\left( C_3\cdot\mathcal{Z}\right).
\end{equation}
The $C$ matrices in the above integrals are given by
\begin{equation}
C_1= 
\begin{pmatrix} 
a_1 & b_1 & c_1 & d_1  & 0 & 0 & 1 \\
0 & 0 & e_1 & f_1 & g_1 & h_1 & 1 
\end{pmatrix},
\label{eq:C1a}
\end{equation}
\begin{equation}
C_2=
\begin{pmatrix} 
a_2 & b_2 & 0 & 0 & c_2 & d_2 & 1 \\
e_2 & f_2 & g_2 & h_2 & 0 & 0 & 1
\end{pmatrix},
\label{eq:C2a}
\end{equation}
\begin{equation}
 C_3=
\begin{pmatrix} 
a_3 & b_3 & 0 & 0 & c_3 & d_3 & 1 \\
0 & 0 & e_3 & f_3 & g_3 & h_3 & 1
\end{pmatrix}.
\label{eq:C3a}
\end{equation}
Each expression clearly has a pole at the point corresponding to the propagator ends coinciding (e.g. $c_1f_1=d_1e_1$ for the first case for example).

The claim is that in the  the sum of the diagrams,  the residues at these poles precisely cancel 
\begin{equation}
\label{eq:ThreeWayLimit}
	\operatorname*{Res}_{c_1f_1=d_1e_1 } \mathcal{I}(D_1) + \operatorname*{Res}_{a_2f_2=b_2e_2 }\mathcal{I}(D_2) + \operatorname*{Res}_{d_3g_3=h_3c_3 } \mathcal{I}(D_3)= 0.
\end{equation}

To see this it is useful to change variables.  Using $\mathcal{I}(D_1)$ as an example, make the following change of variables from $(e_1,f_1)$ to $(\alpha,\epsilon_1)$: $e_1 = \alpha c_1 $ and $f_1 = \alpha d_1 + \epsilon_1$ so that the spurious pole is at  $\epsilon_1=0$.   Substituting these in we have
\begin{align}
\label{eq:C1IntegralLimit}
\operatorname*{Res}_{\epsilon_1=0 }\mathcal{I}(D_1)&= \operatorname*{Res}_{\epsilon_1=0 }\int \frac{\diff a_1 \diff b_1  \diff c_1 \diff d_1 \diff g_1 \diff h_1 \diff \alpha \diff \epsilon_1}{a_1b_1c_1d_1g_1h_1\alpha \epsilon_1}\delta^{(8|8)}\left( C_1\cdot\mathcal{Z}\right)\notag \\
&= \int \frac{\diff a_1 \diff b_1  \diff c_1 \diff d_1 \diff g_1 \diff h_1 \diff \alpha }{a_1b_1c_1d_1g_1h_1\alpha }\delta^{(8|8)}\left( C_1|_{\epsilon_1=0}\cdot\mathcal{Z}\right)
\end{align}
with 
\begin{equation}
\label{eq:C1Limit}
C_1= 
\begin{pmatrix} 
a_1 & b_1 & c_1 & d_1  & 0 & 0 & 1 \\
0 & 0 & \alpha c_1 & \alpha d_1 +\epsilon_1& g_1 & h_1 & 1 
\end{pmatrix},
\end{equation}
The residues  of the other two integrals are dealt with in a similar manner. Changing coordinates from $(e_2,f_2)$ to $\beta,\epsilon_2$ and from $(g_3,h_3)$ to $\gamma,\epsilon_3$   with $e_2=\beta a_2, f_2=\beta b_2+\epsilon_2$ and $g_3=\gamma c_3+\epsilon_3, h_3=\gamma d_3$
the measure then has a simple dlog form in all variables (just as in~\eqref{eq:C1IntegralLimit}. 
Taking the residue at $\epsilon_i \rightarrow 0$ then yields  
\begin{equation}
C_2|_{\epsilon_2=0}=
\begin{pmatrix} 
{a_2} & {b_2} & 0 & 0 & c_2 & d_2 & 1 \\
\beta a_2 & \beta b_2 & g_2 & h_2 & 0 & 0 & 1
\end{pmatrix},
\label{eq:C2}
\end{equation}
\begin{equation}
C_3|_{\epsilon_3=0}=
\begin{pmatrix} 
a_3 & b_3 & 0 & 0 & c_3 & d_3 & 1 \\
0 & 0 & e_3 & f_3 & \gamma c_3 & \gamma d_3 & 1
\end{pmatrix},
\label{eq:C3}
\end{equation}
and  the remaining measure being simply the dlog of all variables as in~\eqref{eq:C1IntegralLimit}.

In order to compare the three $C_i \in \Gr(2,7)$, a change of basis must be introduced for $C_2$ and $C_3$.  Utilising the $GL(2)$ invariance, we define
\begin{equation}
\label{eq:C2Transformation}
C'_2 =
\begin{pmatrix}
0 & 1 \\
\frac{-\beta}{1-\beta} & \frac{1}{1-\beta}
\end{pmatrix} C_2
\end{equation}
and
\begin{equation}
\label{eq:C3Transformation}
C'_3 =
\begin{pmatrix}
\frac{-\gamma}{1-\gamma} & \frac{1}{1-\gamma} \\
0 & 1
\end{pmatrix}C_3.
\end{equation}
The matrices $C'_2$ and $C'_3$ are now of the same form as $C_1$, meaning all three matrices have zeros and ones in the same entries and variables in all of the others: 
\begin{equation}
\label{eq:C2Limit}
C'_2= 
\begin{pmatrix} 
\beta a_2 & \beta b_2 & g_2 & h_2 & 0 & 0 & 1 \\
0 & 0 & \frac{g_2}{1-\beta} & \frac{h_2}{1-\beta} & \frac{-\beta c_2}{1-\beta} & \frac{-\beta d_2}{1-\beta} & 1 
\end{pmatrix},
\end{equation}
\begin{equation}
\label{eq:C3Limit}
C'_3= 
\begin{pmatrix} 
\frac{-\gamma a_3}{1-\gamma} & \frac{-\gamma b_3}{1- \gamma} & \frac{e_3}{1-\gamma} & \frac{f_3}{1-\gamma}  & 0 & 0 & 1 \\
0 & 0 & e_3 & f_3 & \gamma c_3 & \gamma d_3 & 1 
\end{pmatrix}.
\end{equation}  
Each entry of these two matrices can now be compared directly to the equivalent entry in $C_1$. We then change variables again from  $a_2, \ldots , h_2$ and $a_3, \ldots , f_3$ to  $a_1, \ldots, h_1$ as dictated by matching the entries of $C'_2,C'_3$ to those of $C_1$.  In particular we replace  $\beta = \frac{-(1-\alpha)}{\alpha}$ and $\gamma = 1-\alpha$.  Substituting these into the residues of $\mathcal{I}(D_2)$ and $\mathcal{I}(D_3)$, and taking the sum of all three integrals gives
\begin{equation}
\label{eq:IntegralsSummed}
\int \frac{\diff a_1 \diff b_1 \diff c_1 \diff d_1dg_1 \diff h_1 \diff \alpha}{a_1b_1c_1d_1g_1h_1}\left(\frac{1}{\alpha} + \frac{1}{1-\alpha} -\frac{1}{\alpha(1-\alpha)}\right)\delta^{(8|8)}\left( C_1\cdot\mathcal{Z}\right) = 0,
\end{equation}
therefore (\ref{eq:ThreeWayLimit}) is indeed satisfied. 

We now wish to interpret this calculation geometrically.  This cancellation does indeed have a geometric interpretation, as a three-way pairwise matching of the corresponding spurious boundaries. However as we will show there is no way to assign geometries to be consistent with the three way cancellation described above, as well as the other spurious pole cancellations.

\subsection{Spurious boundary matching}

We wish to  associate a geometrical subspace of $\Gr(2,6)$ for each N${}^2$MHV WLD such that the spurious boundaries all match pairwise with those of other diagrams.%
\footnote{Although this may not be a sufficient condition to ensure a good geometry without spurious boundaries, it is necessary.}  
It is straightforward to read off a geometrical region  whose canonical form gives the WLD volume form. In the coordinates we used in the previous section we have a dlog form for the measure (see for example~\eqref{eq:C1IntegralLimit}).
We expect therefore that the corresponding geometry corresponds to simply taking these coordinates, making them real and assigning signs to them. So for example, the diagram in Figure~\ref{fig:threediag}a, using the  coordinates chosen in~\eqref{eq:C1IntegralLimit}, corresponds to a dlog volume form (see the first line of~\eqref{eq:C1IntegralLimit})
  and hence we expect it to be the canonical form of the region
\begin{align}
\begin{array}{c}
\{Y=C_1.Z:   a_1{>}0,\,b_1{>}0,\,d_1{>}0,\,e_1{>}0,\,g_1{>}0,\,h_1{>}0,\,\alpha>0,\,\epsilon_1>0\}\\
=\\
\{Y=C_1.Z:   a_1{>}0,\,b_1{>}0,\,c_1>0,\,d_1{>}0,\,e_1{>}0,\,g_1{>}0,\,h_1{>}0,\,f_1 c_1>e_1 d_1\}
\end{array}		
\end{align}
with $C_1$ given in~\eqref{eq:C1Limit}.
But this is not unique, other  sign choices for the variables   can be chosen to give another region with the same canonical form.%
\footnote{\label{foonote4}There are eight allowed possibilities for the parameters $c_1,d_1,e_1,f_1$ associated with the propagator ends which are on the same edge. These  correspond to choosing signs $s_1,s_2$ for $d_1$ and $e_1$ (four different cases). We then require  $s_1s_2(c_1f_1-e_1d_1>0)$ which splits into two disconnected regions which can be distinguished by the signs of $c_1$ or $f_1$. This gives two possibilities for each of the four cases, or eight cases in total. Very nicely, these cases can also be read off from the parametrisation of the WLD if we think of the parameters as real instead of complex. In order for the ends not to cross we require either $0<d_1/c_1<f_1/e_1$ or $0<e_1/f_1<c_1/d_1$. Then choosing signs for $d_1,e_1$ gives the same eight cases as above.}
 So the challenge is to choose consistent signs so that all spurious boundaries match pairwise.

We begin by looking at the geometric interpretation of the three way cancellation described in the previous section to give some insight.  In order to do this, compare the rotated matrices in the appropriate limit corresponding to the spurious boundary where two propagator ends meet (described in the previous subsection)
\begin{equation}
\label{eq:C1'}
C_1= 
\begin{pmatrix} 
a_1 & b_1 & c_1 & d_1  & 0 & 0 & 1 \\
0 & 0 & \alpha c_1 & \alpha d_1 & g_1 & h_1 & 1 
\end{pmatrix}
\end{equation}
\begin{equation}
\label{eq:C2'}
C'_2= 
\begin{pmatrix} 
\beta a_2 & \beta b_2 & g_2 & h_2  & 0 & 0 & 1 \\
0 & 0 & \frac{g_2}{1-\beta} & \frac{h_2}{1-\beta} & \frac{-\beta c_2}{1-\beta} & \frac{-\beta d_2}{1-\beta} & 1 
\end{pmatrix}
\end{equation}
\begin{equation}
\label{eq:C3'}
C'_3= 
\begin{pmatrix} 
\frac{-\gamma a_3}{1-\gamma} & \frac{-\gamma b_3}{1-\gamma} & \frac{e_3}{1-\gamma} & \frac{f_3}{1-\gamma}  & 0 & 0 & 1 \\
0 & 0 & e_3 & f_3 & \gamma c_3 & \gamma d_3 & 1 
\end{pmatrix}\ .
\end{equation}
At points where the regions touch we thus have
 $\alpha = \frac{1}{1-\beta}$ and $\alpha = 1-\gamma$.  
We now need to choose signs (positive or negative) for the variables $\alpha$, $\beta$ and $\gamma$ such that $\alpha$, $\beta (\alpha )$ and $\gamma (\alpha )$ share boundaries pairwise.  Two different cases arise from this consideration:

\begin{enumerate}
	\item One of the variables is positive and the other two negative.  Without loss of generality  we consider  $\alpha >0$, $ \beta,\gamma < 0$.  
	\item $\alpha$, $\beta$ and $\gamma$ are all positive.
\end{enumerate}

\begin{figure}[H]
	\begin{center}
		\includegraphics[scale=0.5]{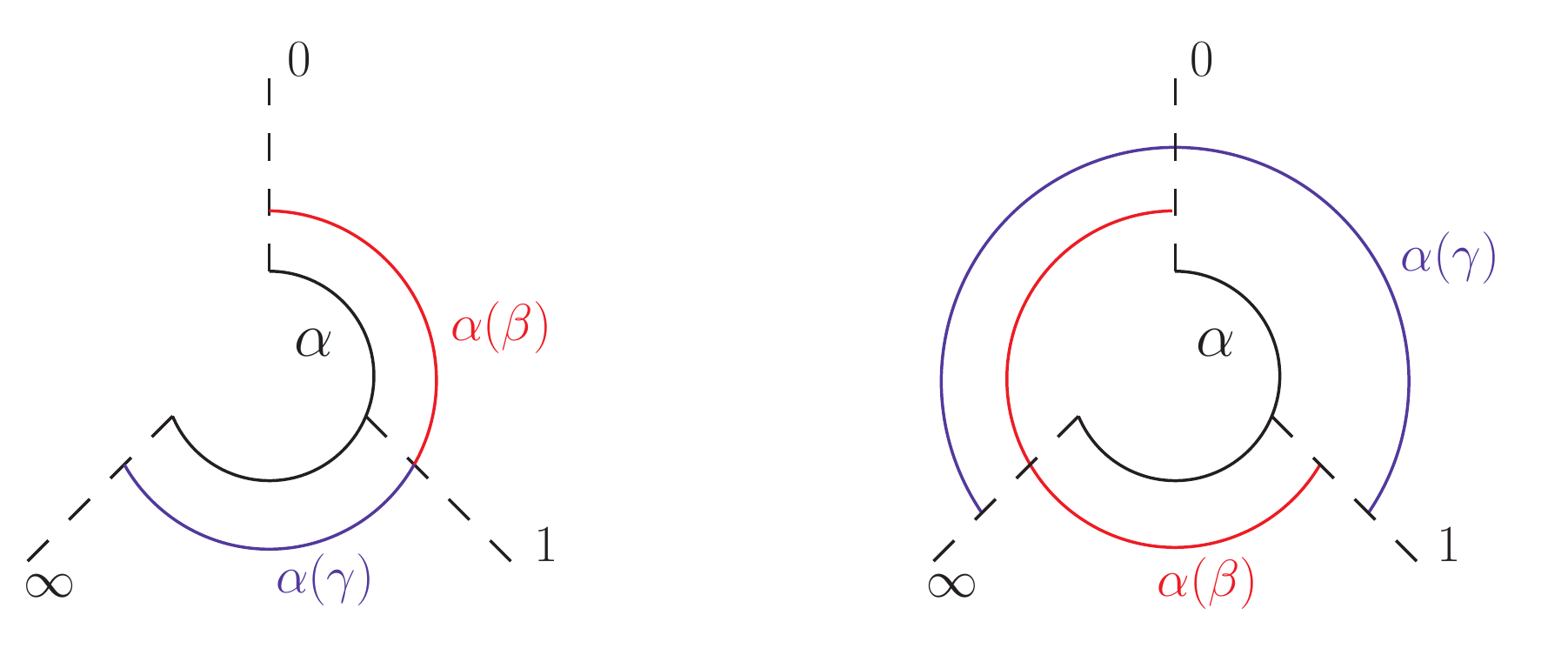}
		\caption{The two possibilities for three way boundary matching. We plot the range of $\alpha$ on a circle from $[-\infty,\infty]$ passing through 0 and 1. Black is the range of $\alpha$ in diagram D1, in red that of $\alpha(\beta)$ in $D_2$ and in blue the range of   $\alpha(\gamma)$ in $D_3$. We see there is always a pairwise matching of the three diagrams in both cases. In Case 1 D2 and D3 each only overlap with D1 and not with each other. For Case 2 all diagrams overlap the other two.}
		\label{fig:AlphaSpace}
	\end{center}
\end{figure}

The two cases are illustrated in Figure~\ref{fig:AlphaSpace}.

\subsubsection{Case 1: $\alpha >0$, $\beta <0$ and $\gamma <0$}
Looking at Figure~\ref{fig:AlphaSpace}a, $C_1$ and $C'_2$ should overlap when $0<\alpha<1$ whereas $C_1$ and $C'_3$ should overlap when $1<\alpha<\infty$.  

Now at the points where the regions overlap we also need all other variables to match. In particular this fixes the signs of the variables for the second two diagrams in terms of the first. Defining
\begin{align}
	&\sgn(a_1)=s_1,\quad \sgn(b_1)=s_2,\quad \sgn(c_1)=s_3,\notag \\ &\sgn(d_1)=s_4,\quad \sgn(g_1)=s_5,\quad \sgn(h_1)=s_6\label{signs}
\end{align} 
Then by comparing~(\ref{eq:C2'},\ref{eq:C3'}) to~\eqref{eq:C1'} and undoing the $GL(2)$ transformation we must have the following signs in each entry 
\begin{equation}
\label{eq:C1SignedCase1}
\sgn(C_1)= 
\begin{pmatrix} 
s_1 & s_2 & s_3 & s_4  & 0 & 0 & 1 \\
0 & 0 & s_3 & s_4 & s_5 & s_6 & 1 
\end{pmatrix}
\end{equation}
\begin{equation}
\label{eq:C2SignedCase1}
\sgn(C_2)= 
\begin{pmatrix} 
-s_1 & -s_2 & 0 & 0  & s_5 & s_6 & 1 \\
s_1 & s_2 & s_3 & s_4 & 0 & 0 & 1 
\end{pmatrix}
\end{equation}
\begin{equation}
\label{eq:C3SignedCase1}
\sgn(C_3)= 
\begin{pmatrix} 
s_1 & s_2 & 0 & 0 & -s_5 & -s_6 & 1 \\
0 & 0 & s_3 & s_4 & s_5 & s_6 & 1 
\end{pmatrix}.
\end{equation}
Given a set of signs for $C_1$, the three way cancellation fixes the signs of $C_2$ and $C_3$.
Although these signs are derived by looking at their values at the spurious boundary, crucially the signs remain unchanged inside the region even on moving away from the boundary.%
\footnote{The only possible exception to this would be those entries depending on $\epsilon_i$.  For example the entry $\alpha d_1 +\epsilon_1$ in~\eqref{eq:C1Limit} if $\epsilon_1$ were to have a different sign to $\alpha d_1$. This corresponds to one of the  disallowed possibilities (see footnote~\ref{foonote4}). In any case all entries of any $C$ matrix do need to have definite signs to match spurious boundaries of nearby diagrams where the propagators end on different edges.}

But now the sign choices for diagrams $D_1,D_2,D_3$ (forced on us by the three way cancellation Case 1) can be seen to be inconsistent with the consecutive matching of the other type of spurious boundary where propagator ends approach vertices. The problem comes down to  the difference in signs in the top row of~\eqref{eq:C2SignedCase1} with those of~\eqref{eq:C3SignedCase1}.

Now, consider starting with diagram $D_2$ and moving the propagator defined by the second line in $C_2$ around clockwise until the diagram $D_3$ is reached.  At each vertex we match spurious boundaries, meaning the signs of the top row (corresponding to the propagator left fixed) must remain the same. Under this sequence of moves
\begin{equation}
\label{eq:C2Moved}
\sgn(C_2) \to
\begin{pmatrix} 
-s_1 & -s_2 & 0 & 0 & s_5 & s_6 & 1 \\
0 & 0 & s'_3 & s'_4 & s'_5 & s'_6 & 1 
\end{pmatrix},
\end{equation}
where the prime variables represent new signs not fixed in this process.%
\footnote{In fact we require  $s_3'=s_3$ and $s_4'=s_4$ by the same argument as for the NMHV case: consecutive spurious boundaries implies fixed signs per vertex for a propagator end, see discussion around~\eqref{NMHVtes2}.}

Now  comparing this new matrix to $C_3$~\eqref{eq:C3SignedCase1} one can see immediately that the signs on the top row are different, regardless of what the bottom row becomes.  Therefore, the signs that are found from the matching of the three-way spurious boundary are not consistent with the matching of boundaries obtained by following the propagators round the Wilson Loop polygon.  The WLDs cannot be glued together to form a geometry without spurious boundaries with this choice of $\alpha$, $\beta$ and $\gamma$.

Note this argument has been illustrated for at points but clearly doesn't depend in any key way on the number of points.

\subsubsection{Case 2: $\alpha,\beta,\gamma > 0$}

We then consider  the second possibility for having pairwise matching boundaries where $\alpha,\beta,\gamma > 0$.

Looking at Figure~\ref{fig:AlphaSpace}b, $C_1$ and $C'_2$ should overlap when $1<\alpha<\infty$ and $0<\beta<1$ and $C_1$ and $C'_3$ should overlap when $0<\alpha<1$ and $0<\gamma<1$.  Now there is an additional overlap between $C'_2$ and $C'_3$ when $1<\beta<\infty$ and $1<\gamma<\infty$.  

At these overlaps the entries of the rotated matrices~(\ref{eq:C1'}-\ref{eq:C3'}) must be equal. Defining the signs of the $C_1$ variables as previously~\eqref{signs} this means the signs of the entries of $C_2'$ and $C_3'$, must be the same as those of $C_1$ in the region where they overlap with $C_1$ (i.e.  $0<\beta<1$, $0<\gamma<1$). However when $\beta,\gamma>1$ some of the entries changes sign due their dependence on $\beta$ or $\gamma$. Thus the signs of the entries of the rotated $C$ matrices are as follows:
\begin{equation}
\label{eq:C1'SignedCase2_2}
\sgn(C_1): 
\begin{pmatrix} 
s_1 & s_2 & s_3 & s_4  & 0 & 0 & 1 \\
0 & 0 & s_3 & s_4 & s_5 & s_6 & 1 
\end{pmatrix}_{0<\alpha<\infty}
\end{equation}
\begin{equation}
\label{eq:C2'SignedCase2_2}
\sgn(C'_2): 
\begin{pmatrix} 
s_1 & s_2 & s_3 & s_4 & 0 & 0 & 1 \\
0 & 0 & s_3 & s_4 & s_5 & s_6 & 1 
\end{pmatrix}_{0<\beta<1},
\begin{pmatrix} 
s_1 & s_2 & s_3 & s_4 & 0 & 0 & 1 \\
0 & 0 & -s_3 & -s_4 & -s_5 & -s_6 & 1
\end{pmatrix}_{1<\beta<\infty}
\end{equation}
\begin{equation}
\label{eq:C3'SignedCase2_2}
\sgn(C'_3): 
\begin{pmatrix} 
s_1 & s_2 & s_3 & s_4 & 0 & 0 & 1 \\
0 & 0 & s_3 & s_4 & s_5 & s_6 & 1 
\end{pmatrix}_{0<\gamma<1},
\begin{pmatrix} 
-s_1 & -s_2 & -s_3 & -s_4 & 0 & 0 & 1 \\
0 & 0 & s_3 & s_4 & s_5 & s_6 & 1 
\end{pmatrix}_{1<\gamma<\infty}.
\end{equation}
But now there is a clear problem. 
Looking at the matrices for $1<\beta<\infty$ and $1<\gamma<\infty$, it can be seen they do not match as they should. 
Matching diagram $D_2$ with $D_1$ correctly and $D_3$ with $D_1$ correctly fixes the signs of $D_2,D_3$ in a way incompatible with $D_2$ and $D_3$ matching.

Thus there is in fact no valid three way boundary matching for this case.

\section{Conclusion}

We have shown that surprisingly it is not possible to  consistently assign a subspace of $\Gr(k,k+4)$ (amplituhedron space) to each WLD consistent with its canonical form and pairwise matching  of all spurious boundaries. 
In other words  WLDs can not be used to tessellate the amplituhedron or any other shape without spurious boundaries.
This despite their promising properties: WLDs do have natural  (but non-unique) interpretations as subspaces in $\Gr(k,k+4)$ and they do sum up to give the amplitude. The situation is similar to the example in Figure~\ref{P2examples}b where we see an attempted  tessellation of the quadrilateral: although the canonical forms of the two triangles sum to the corresponding canonical forms of the quadrilateral, this is clearly not a tessellation of the quadrilateral and there are left over unmatched spurious boundaries. Of course for the quadrilateral we could choose a more sensible tessellation with matched spurious boundaries, for WLDs we have shown there is no such sensible tessellation possible.

Note that we have shown this for the N${}^2$MHV case and illustrated for six points only. We have already mentioned that the proof does not depend on the number of points.  It is also clear that the proof goes through  in the same way for higher MHV degree:  just add another propagator somewhere away from the three way cancellation and recycle the same argument given here.
We have also here focussed on tree level but it would be very surprising if moving to loop level improves the situation.

One might hope that although the WLDs do not tessellate the amplituhedron they may instead give a nice tessellation of the squared amplituhedron~\cite{Eden:2017fow,Arkani-Hamed:2017vfh}
which has a  more direct definition and for which there are $2^k$ copies of most diagrams, which could conceivably provide a way out of the problems found here. However this also seems not to be the case (although the proof is more involved and we omit it here).

We should emphasise that despite the fact that the WLDs can not provide a geometric tessellation of the amplituhedron, they do still give a very concrete and suggestive ``tessellation'' at the level of its canonical form. It seems likely that this property  generalises for more general positive Grassmannians and may prove useful in their further mathematical study.

\subsection*{Acknowledgements}

PH would like to thank Lionel Mason for many invaluable discussions on topics related to this. PH and AS would also like to thank Susama Agarwala and other members of the WLD virtual seminar series for the interesting seminars and discussions. 
AS is supported by an STFC studentship and PH acknowledges support from STFC grant ST/P000371/1.


\begin{thebibliography}{99}
%\cite{0905.1473}
\bibitem{0905.1473}
A.~Hodges,
``Eliminating spurious poles from gauge-theoretic amplitudes,''
JHEP {\bf 1305} (2013) 135
doi:10.1007/JHEP05(2013)135
[arXiv:0905.1473 [hep-th]].
%%CITATION = doi:10.1007/JHEP05(2013)135;%%
%231 citations counted in INSPIRE as of 12 Jul 2018


%\cite{1012.6030}
\bibitem{1012.6030}
N.~Arkani-Hamed, J.~L.~Bourjaily, F.~Cachazo, A.~Hodges and J.~Trnka,
``A Note on Polytopes for Scattering Amplitudes,''
JHEP {\bf 1204} (2012) 081
doi:10.1007/JHEP04(2012)081
[arXiv:1012.6030 [hep-th]].
%%CITATION = doi:10.1007/JHEP04(2012)081;%%
%54 citations counted in INSPIRE as of 12 Jul 2018


%\cite{1312.2007}
\bibitem{1312.2007}
N.~Arkani-Hamed and J.~Trnka,
``The Amplituhedron,''
JHEP {\bf 1410} (2014) 030
doi:10.1007/JHEP10(2014)030
[arXiv:1312.2007 [hep-th]].
%%CITATION = doi:10.1007/JHEP10(2014)030;%%
%173 citations counted in INSPIRE as of 12 Jul 2018


%\cite{Arkani-Hamed:2013kca}
\bibitem{Arkani-Hamed:2013kca}
N.~Arkani-Hamed and J.~Trnka,
``Into the Amplituhedron,''
JHEP {\bf 1412} (2014) 182
doi:10.1007/JHEP12(2014)182
[arXiv:1312.7878 [hep-th]].
%%CITATION = doi:10.1007/JHEP12(2014)182;%%
%84 citations counted in INSPIRE as of 12 Jul 2018


%\cite{Bai:2014cna}
\bibitem{Bai:2014cna}
Y.~Bai and S.~He,
``The Amplituhedron from Momentum Twistor Diagrams,''
JHEP {\bf 1502} (2015) 065
doi:10.1007/JHEP02(2015)065
[arXiv:1408.2459 [hep-th]].
%%CITATION = doi:10.1007/JHEP02(2015)065;%%
%40 citations counted in INSPIRE as of 12 Jul 2018


%\cite{Franco:2014csa}
\bibitem{Franco:2014csa}
S.~Franco, D.~Galloni, A.~Mariotti and J.~Trnka,
``Anatomy of the Amplituhedron,''
JHEP {\bf 1503} (2015) 128
doi:10.1007/JHEP03(2015)128
[arXiv:1408.3410 [hep-th]].
%%CITATION = doi:10.1007/JHEP03(2015)128;%%
%33 citations counted in INSPIRE as of 12 Jul 2018


%\cite{Lam:2014jda}
\bibitem{Lam:2014jda}
T.~Lam,
``Amplituhedron cells and Stanley symmetric functions,''
Commun.\ Math.\ Phys.\  {\bf 343} (2016) no.3,  1025
doi:10.1007/s00220-016-2602-2
[arXiv:1408.5531 [math.AG]].
%%CITATION = doi:10.1007/s00220-016-2602-2;%%
%17 citations counted in INSPIRE as of 12 Jul 2018


%\cite{Arkani-Hamed:2014dca}
\bibitem{Arkani-Hamed:2014dca}
N.~Arkani-Hamed, A.~Hodges and J.~Trnka,
``Positive Amplitudes In The Amplituhedron,''
JHEP {\bf 1508} (2015) 030
doi:10.1007/JHEP08(2015)030
[arXiv:1412.8478 [hep-th]].
%%CITATION = doi:10.1007/JHEP08(2015)030;%%
%33 citations counted in INSPIRE as of 12 Jul 2018

%\cite{Agarwala:2015vma}
\bibitem{Agarwala:2015vma}
S.~Agarwala and E.~Marin-Amat,
%``Wilson Loop diagrams and Positroids,''
Commun.\ Math.\ Phys.\  {\bf 350} (2017) no.2,  569
doi:10.1007/s00220-016-2659-y
[arXiv:1509.06150 [math-ph]].
%%CITATION = doi:10.1007/s00220-016-2659-y;%%
%1 citations counted in INSPIRE as of 16 Jul 2018

%\cite{Bai:2015qoa}
\bibitem{Bai:2015qoa}
Y.~Bai, S.~He and T.~Lam,
``The Amplituhedron and the One-loop Grassmannian Measure,''
JHEP {\bf 1601} (2016) 112
doi:10.1007/JHEP01(2016)112
[arXiv:1510.03553 [hep-th]].
%%CITATION = doi:10.1007/JHEP01(2016)112;%%
%14 citations counted in INSPIRE as of 12 Jul 2018


%\cite{Ferro:2015grk}
\bibitem{Ferro:2015grk}
L.~Ferro, T.~Lukowski, A.~Orta and M.~Parisi,
``Towards the Amplituhedron Volume,''
JHEP {\bf 1603} (2016) 014
doi:10.1007/JHEP03(2016)014
[arXiv:1512.04954 [hep-th]].
%%CITATION = doi:10.1007/JHEP03(2016)014;%%
%22 citations counted in INSPIRE as of 12 Jul 2018


%\cite{Bern:2015ple}
\bibitem{Bern:2015ple}
Z.~Bern, E.~Herrmann, S.~Litsey, J.~Stankowicz and J.~Trnka,
``Evidence for a Nonplanar Amplituhedron,''
JHEP {\bf 1606} (2016) 098
doi:10.1007/JHEP06(2016)098
[arXiv:1512.08591 [hep-th]].
%%CITATION = doi:10.1007/JHEP06(2016)098;%%
%41 citations counted in INSPIRE as of 12 Jul 2018


%\cite{Galloni:2016iuj}
\bibitem{Galloni:2016iuj}
D.~Galloni,
``Positivity Sectors and the Amplituhedron,''
arXiv:1601.02639 [hep-th].
%%CITATION = ARXIV:1601.02639;%%
%12 citations counted in INSPIRE as of 12 Jul 2018


%\cite{Karp:2016uax}
\bibitem{Karp:2016uax}
S.~N.~Karp and L.~K.~Williams,
``The m=1 amplituhedron and cyclic hyperplane arrangements,''
arXiv:1608.08288 [math.CO].
%%CITATION = ARXIV:1608.08288;%%
%7 citations counted in INSPIRE as of 12 Jul 2018


%\cite{Dennen:2016mdk}
\bibitem{Dennen:2016mdk}
T.~Dennen, I.~Prlina, M.~Spradlin, S.~Stanojevic and A.~Volovich,
``Landau Singularities from the Amplituhedron,''
JHEP {\bf 1706} (2017) 152
doi:10.1007/JHEP06(2017)152
[arXiv:1612.02708 [hep-th]].
%%CITATION = doi:10.1007/JHEP06(2017)152;%%
%10 citations counted in INSPIRE as of 12 Jul 2018


%\cite{Ferro:2016zmx}
\bibitem{Ferro:2016zmx}
L.~Ferro, T.~Lukowski, A.~Orta and M.~Parisi,
``Yangian symmetry for the tree amplituhedron,''
J.\ Phys.\ A {\bf 50} (2017) no.29,  294005
doi:10.1088/1751-8121/aa7594
[arXiv:1612.04378 [hep-th]].
%%CITATION = doi:10.1088/1751-8121/aa7594;%%
%8 citations counted in INSPIRE as of 12 Jul 2018


%\cite{Ferro:2016ptt}
\bibitem{Ferro:2016ptt}
L.~Ferro, T.~Lukowski, A.~Orta and M.~Parisi,
``Tree-level scattering amplitudes from the amplituhedron,''
J.\ Phys.\ Conf.\ Ser.\  {\bf 841} (2017) no.1,  012037
doi:10.1088/1742-6596/841/1/012037
[arXiv:1612.06276 [hep-th]].
%%CITATION = doi:10.1088/1742-6596/841/1/012037;%%
%2 citations counted in INSPIRE as of 12 Jul 2018


%\cite{Eden:2017fow}
\bibitem{Eden:2017fow}
B.~Eden, P.~Heslop and L.~Mason,
``The Correlahedron,''
JHEP {\bf 1709} (2017) 156
doi:10.1007/JHEP09(2017)156
[arXiv:1701.00453 [hep-th]].
%%CITATION = doi:10.1007/JHEP09(2017)156;%%
%8 citations counted in INSPIRE as of 12 Jul 2018


%\cite{Arkani-Hamed:2017vfh}
\bibitem{Arkani-Hamed:2017vfh}
N.~Arkani-Hamed, H.~Thomas and J.~Trnka,
``Unwinding the Amplituhedron in Binary,''
JHEP {\bf 1801} (2018) 016
doi:10.1007/JHEP01(2018)016
[arXiv:1704.05069 [hep-th]].
%%CITATION = doi:10.1007/JHEP01(2018)016;%%
%14 citations counted in INSPIRE as of 12 Jul 2018


%\cite{Karp:2017ouj}
\bibitem{Karp:2017ouj}
S.~N.~Karp, L.~K.~Williams and Y.~X.~Zhang,
%``Decompositions of amplituhedra,''
arXiv:1708.09525 [math.CO].
%%CITATION = ARXIV:1708.09525;%%
%4 citations counted in INSPIRE as of 13 Jul 2018



%\cite{Rao:2017fqc}
\bibitem{Rao:2017fqc}
J.~Rao,
``4-particle Amplituhedron at 3-loop and its Mondrian Diagrammatic Implication,''
JHEP {\bf 1806} (2018) 038
doi:10.1007/JHEP06(2018)038
[arXiv:1712.09990 [hep-th]].
%%CITATION = doi:10.1007/JHEP06(2018)038;%%
%2 citations counted in INSPIRE as of 12 Jul 2018


%\cite{An:2017tbf}
\bibitem{An:2017tbf}
Y.~An, Y.~Li, Z.~Li and J.~Rao,
``All-loop Mondrian Diagrammatics and 4-particle Amplituhedron,''
JHEP {\bf 1806} (2018) 023
doi:10.1007/JHEP06(2018)023
[arXiv:1712.09994 [hep-th]].
%%CITATION = doi:10.1007/JHEP06(2018)023;%%
%2 citations counted in INSPIRE as of 12 Jul 2018


%\cite{Arkani-Hamed:2017fdk}
\bibitem{Arkani-Hamed:2017fdk}
N.~Arkani-Hamed, P.~Benincasa and A.~Postnikov,
``Cosmological Polytopes and the Wavefunction of the Universe,''
arXiv:1709.02813 [hep-th].
%%CITATION = ARXIV:1709.02813;%%
%6 citations counted in INSPIRE as of 12 Jul 2018


%\cite{Arkani-Hamed:2017mur}
\bibitem{Arkani-Hamed:2017mur}
N.~Arkani-Hamed, Y.~Bai, S.~He and G.~Yan,
``Scattering Forms and the Positive Geometry of Kinematics, Color and the Worldsheet,''
JHEP {\bf 1805} (2018) 096
doi:10.1007/JHEP05(2018)096
[arXiv:1711.09102 [hep-th]].
%%CITATION = doi:10.1007/JHEP05(2018)096;%%
%11 citations counted in INSPIRE as of 12 Jul 2018


%\cite{Galashin:2018fri}
\bibitem{Galashin:2018fri}
P.~Galashin and T.~Lam,
``Parity duality for the amplituhedron,''
arXiv:1805.00600 [math.CO].
%%CITATION = ARXIV:1805.00600;%%
%1 citations counted in INSPIRE as of 12 Jul 2018


%\cite{Agarwala:2018fms}
\bibitem{Agarwala:2018fms}
S.~Agarwala and S.~Fryer,
``A study in $\mathbb{G}_{\mathbb{R}, \geq 0}$: from the geometric case book of Wilson loop diagrams and SYM $N=4$,''
arXiv:1803.00958 [math.CO].
%%CITATION = ARXIV:1803.00958;%%


%\cite{Ferro:2018vpf}
\bibitem{Ferro:2018vpf}
L.~Ferro, T.~Lukowski and M.~Parisi,
``Amplituhedron meets Jeffrey-Kirwan Residue,''
arXiv:1805.01301 [hep-th].
%%CITATION = ARXIV:1805.01301;%%


%\cite{Rao:2018uta}
\bibitem{Rao:2018uta}
J.~Rao,
``4-particle Amplituhedronics for 3-5 loops,''
arXiv:1806.01765 [hep-th].
%%CITATION = ARXIV:1806.01765;%%


%\cite{Bourjaily:2018bbb}
\bibitem{Bourjaily:2018bbb}
J.~Bourjaily and H.~Thomas,
``What is the Amplituhedron?,''
Not.\ Amer.\ Math.\ Soc.\  {\bf 65} (2018) no.2,  167.
doi:10.1090/noti1630
%%CITATION = doi:10.1090/noti1630;%%




%\cite{1703.04541}
\bibitem{1703.04541}
N.~Arkani-Hamed, Y.~Bai and T.~Lam,
``Positive Geometries and Canonical Forms,''
JHEP {\bf 1711} (2017) 039
doi:10.1007/JHEP11(2017)039
[arXiv:1703.04541 [hep-th]].
%%CITATION = doi:10.1007/JHEP11(2017)039;%%
%15 citations counted in INSPIRE as of 12 Jul 2018

%\cite{1807.05397}
\bibitem{1807.05397}
S.~Agarwala and C.~Marcott,
``Wilson loops in SYM N=4 do not  parametrize an orientable space,''
arXiv:1807.05397 [math-ph].
%%CITATION = ARXIV:1807.05397;%%


%\cite{Mason:2010yk}
\bibitem{Mason:2010yk}
L.~J.~Mason and D.~Skinner,
``The Complete Planar S-matrix of N=4 SYM as a Wilson Loop in Twistor Space,''
JHEP {\bf 1012} (2010) 018
doi:10.1007/JHEP12(2010)018
[arXiv:1009.2225 [hep-th]].
%%CITATION = doi:10.1007/JHEP12(2010)018;%%
%175 citations counted in INSPIRE as of 12 Jul 2018

%\cite{Chicherin:2016fac}
\bibitem{Chicherin:2016fac}
D.~Chicherin and E.~Sokatchev,
%``$ \mathcal{N} $ = 4 super-Yang-Mills in LHC superspace part I: classical and quantum theory,''
JHEP {\bf 1702} (2017) 062
doi:10.1007/JHEP02(2017)062
[arXiv:1601.06803 [hep-th]].
%%CITATION = doi:10.1007/JHEP02(2017)062;%%
%19 citations counted in INSPIRE as of 27 Jul 2018

%\cite{Chicherin:2016ybl}
\bibitem{Chicherin:2016ybl}
D.~Chicherin, P.~Heslop, G.~P.~Korchemsky and E.~Sokatchev,
``Wilson Loop Form Factors: A New Duality,''
JHEP {\bf 1804} (2018) 029
doi:10.1007/JHEP04(2018)029
[arXiv:1612.05197 [hep-th]].
%%CITATION = doi:10.1007/JHEP04(2018)029;%%
%4 citations counted in INSPIRE as of 12 Jul 2018


%\cite{Alday:2007hr}
\bibitem{Alday:2007hr}
L.~F.~Alday and J.~M.~Maldacena,
``Gluon scattering amplitudes at strong coupling,''
JHEP {\bf 0706} (2007) 064
doi:10.1088/1126-6708/2007/06/064
[arXiv:0705.0303 [hep-th]].
%%CITATION = doi:10.1088/1126-6708/2007/06/064;%%
%729 citations counted in INSPIRE as of 12 Jul 2018


%\cite{Drummond:2007aua}
\bibitem{Drummond:2007aua}
J.~M.~Drummond, G.~P.~Korchemsky and E.~Sokatchev,
``Conformal properties of four-gluon planar amplitudes and Wilson loops,''
Nucl.\ Phys.\ B {\bf 795} (2008) 385
doi:10.1016/j.nuclphysb.2007.11.041
[arXiv:0707.0243 [hep-th]].
%%CITATION = doi:10.1016/j.nuclphysb.2007.11.041;%%
%446 citations counted in INSPIRE as of 12 Jul 2018


%\cite{Brandhuber:2007yx}
\bibitem{Brandhuber:2007yx}
A.~Brandhuber, P.~Heslop and G.~Travaglini,
``MHV amplitudes in N=4 super Yang-Mills and Wilson loops,''
Nucl.\ Phys.\ B {\bf 794} (2008) 231
doi:10.1016/j.nuclphysb.2007.11.002
[arXiv:0707.1153 [hep-th]].
%%CITATION = doi:10.1016/j.nuclphysb.2007.11.002;%%
%401 citations counted in INSPIRE as of 12 Jul 2018


%\cite{Chicherin:2014uca}
\bibitem{Chicherin:2014uca}
D.~Chicherin, R.~Doobary, B.~Eden, P.~Heslop, G.~P.~Korchemsky, L.~Mason and E.~Sokatchev,
``Correlation functions of the chiral stress-tensor multiplet in $ \mathcal{N}=4 $ SYM,''
JHEP {\bf 1506} (2015) 198
doi:10.1007/JHEP06(2015)198
[arXiv:1412.8718 [hep-th]].
%%CITATION = doi:10.1007/JHEP06(2015)198;%%
%20 citations counted in INSPIRE as of 12 Jul 2018




\end{thebibliography}
\end{document}